\documentclass[twocolumn]{aa}
\usepackage{graphicx}
\usepackage{amssymb}
\usepackage{epstopdf}
\usepackage{booktabs}
\usepackage{upgreek}
\usepackage{lscape}
\usepackage[locale=US]{siunitx}
\usepackage{hyperref}

\begin{document}

\title{Multiple origins for the DLA at $z_\mathrm{abs}=0.313$ toward
PKS\,1127$-$145 indicated by a complex dust depletion pattern of 
Ca, Ti, and Mn}
 
\author{
C.~R. Guber \inst{1},
P. Richter \inst{1,2},
\and
M. Wendt \inst{1}
}

\offprints{C.~R. Guber\\
\email{guber@astro.physik.uni-potsdam.de}}

\institute{Institut f\"ur Physik und Astronomie, Universit\"at Potsdam,
           Karl-Liebknecht-Str.\,24/25, 14476 Golm, Germany
\and
Leibniz-Institut f\"ur Astrophysik Potsdam (AIP), An der Sternwarte 16,
14482 Potsdam, Germany
}

\date{Received xxx; accepted xxx}

%%%%%%%%%%%%%%%%%%%%%%%%%%%%%%%%%%%%%%%%%%%%%%%%%%%%%%%%%%%%%%%%%%%%%%%%%%%%%
%%%%%%%%%%%%%%%%%%%%%%%%%%%%%%%%%%%%%%%%%%%%%%%%%%%%%%%%%%%%%%%%%%%%%%%%%%%%%

% \abstract{}{}{}{}{} 
% 5 {} token are mandatory
 
\abstract
{}
{We aim to investigate the dust depletion properties of optically thick gas in and 
around galaxies and its origin we study in detail the dust depletion patterns 
of Ti, Mn, and Ca in the multi-component damped Lyman$\upalpha$ (DLA) absorber 
at $z_\mathrm{abs}=0.313$ toward the quasar PKS\,1127$-$145.}
{We performed a detailed spectral analysis of the absorption profiles of 
Ca\,{\sc ii}, Mn\,{\sc ii}, Ti\,{\sc ii}, and Na\,{\sc i} associated with 
the DLA toward PKS\,1127$-$145, based on optical high-resolution data 
obtained with the UVES instrument at the Very Large Telescope (VLT). 
We obtained column densities and Doppler-parameters for the ions listed 
above and determine their gas-phase abundances, from which we conclude 
on their dust depletion properties. We compared the Ca and Ti depletion 
properties of this DLA with that of other DLAs.}
{One of the six analyzed absorption components (component 3) shows a 
striking underabundance of Ti and Mn in the gas-phase, indicating the 
effect of dust depletion for these elements and a locally enhanced 
dust-to-gas ratio. In this DLA and in other similar absorbers, the 
Mn\,{\sc ii} abundance follows that of Ti\,{\sc ii} very closely, 
implying that both ions are equally sensitive to the dust depletion 
effects.}      
{Our analysis indicates that the DLA toward PKS\,1127$-$145 has multiple 
origins. With its narrow line width and its strong dust depletion, 
component 3 points toward the presence of a neutral gas disk from a faint 
LSB galaxy in front of PKS\,1127$-$145, while the other, more diffuse 
and dust-poor, absorption components possibly are related to tidal gas 
features from the interaction between the various, optically confirmed 
galaxy-group members. In general, the Mn/Ca\,{\sc ii} ratio in sub-DLAs 
and DLAs possibly serves as an important indicator to discriminate between 
dust-rich and dust-poor in neutral gas in and around galaxies.}


\titlerunning{The multicomponent DLA at $z_\mathrm{abs}=0.313$ 
toward PKS\,1127$-$145}

\authorrunning{Guber, Richter, \& Wendt}

\maketitle

%%%%%%%%%%%%%%%%%%%%%%%%%%%%%%%%%%%%%%%%%%%%%%%%%%%%%%%%%%%%%%%%%%%%%%%%%%%%%
%%%%%%%%%%%%%%%%%%%%%%%%%%%%%%%%%%%%%%%%%%%%%%%%%%%%%%%%%%%%%%%%%%%%%%%%%%%%%

\section{Introduction}

Quasar absorption spectroscopy is a powerful method to study diffuse
gas inside and outside of galaxies in the local Universe as well as 
at high redshift. This method provides accurate information 
about the distribution, the chemical composition, and the kinematics 
of the interstellar, 
circumgalactic, and intergalactic medium at low and high redshift
(ISM, CGM, IGM, respectively; 
e.g., \citealt{bahcall69}; \citealt{bergeron86};
\citet{morris93}; \citet{savage2002}).

Most of the diagnostic transitions for quasar (or QSO) absorption 
spectroscopy from the most abundant atoms and ions (e.g., H\,{\sc i}, 
O\,{\sc i}, C\,{\sc iv}, Si\,{\sc iii}, Mg\,{\sc ii}) 
are located in the ultraviolet (UV) at 
wavelengths $<2000$\,\AA\,(see \citealt{bmorton}). The observation 
of these transitions at low redshift
requires a space-based UV spectrograph, such as the Cosmic Origins
Spectrograph (COS) on board the {\it Hubble Space Telescope} (HST).
However, a few strong transitions 
from, if at all, only few times ionized ions of relatively abundant 
elements (e.g., Ca\,{\sc ii},
Ti\,{\sc ii}, Mn\,{\sc ii}, Na\,{\sc i}) are present also in the 
near-UV and optical regime. 
In the low-$z$ Universe, these transitions can be observed using 
high-resolution ground-based spectrographs at 8\,m-class telescopes
(e.g., the UVES instrument at the Very Large Telescope, VLT).

Intervening QSO absorbers with the highest neutral gas column densities
are referred to as damped Lyman$\upalpha$ absorbers 
(DLAs; log $N$(H\,{\sc i}$)\geq 20.3$)
and sub-DLAs ($19.0\leq$\,log\,$N$(H\,{\sc i}$)<20.3$). These systems
are believed to arise in a broad range of galactic and circumgalactic 
environments (e.g., neutral gas disks of late-type galaxies, 
interstellar gas in dwarf galaxies,
cold gas in accretion flows, tidal gas streams from interacting galaxies).
Disentangling these different origins for individual systems requires 
supplementary imaging data (to study the local galaxy environment) and 
(ideally) some other kind of information (e.g., from radio and infrared 
observations).

%While previous absorption-line studies of DLAs and sub-DLAs have provided a wealth 
%of information on gas-phase abundances and absorber kinematics, relatively little 
%attention has been paid to the overall dust properties in the absorbers and intrinsic 
%dust depletion patterns that can be derived from high-resolution spectra. 
%In our previous paper, \citet{bguber}, we have carried out a systematic study of 
%dust depletion in DLAs and sub-DLAs using the Ca\,{\sc ii} and Ti\,{\sc ii} transitions
%in the near-UV/optical regime.

Previous absorption-line studies of DLAs and sub-DLAs have provided 
a wealth of information on gas-phase abundances (e.g., \citealt{bdessauges06}), 
absorber kinematics (e.g., \citealt{brao11}), overall dust properties in the 
absorbers (e.g., \citealt{bdecia13}), and intrinsic dust depletion patterns 
(e.g., \citealt{bdecia}; \citealt{bledoux02}; \citealt{bledoux03}) that can be 
derived from high-resolution spectra.
However, although there are also studies concerning Ca\,{\sc ii} and Ti\,{\sc ii} 
absorption (e.g., \citealt{bledoux02}; \citealt{brichter}; \citealt{bwelty}), 
little attention has been paid to the ratio of these two extremely dust-affine 
elements. 
In our previous paper, \citet{bguber}, we have carried out a systematic study of 
dust depletion in DLAs and sub-DLAs using the Ca\,{\sc ii} and Ti\,{\sc ii} 
transitions and the ratio of these ions in the near-UV and optical regime.

According to each of their ionization potentials in predominantly neutral gas, 
Ti and Mn are particularly present in the form of their one-times ionized 
ions Ti\,{\sc ii} and Mn\,{\sc ii}, measured by us.
Because the ionization potential of Ca\,{\sc ii} is slightly smaller than 
that of H\,{\sc i}, also the second times ionized Ca-ion (Ca\,{\sc iii}) 
may contribute remarkably to the total Ca abundance in such neutral 
environments (e.g., \citealt{bguber}).
Unfortunately, Ca\,{\sc iii} is not spectroscopically measurable in our study. 
To express the fact that we know and use only the Ca\,{\sc ii} column densities 
for calculations, and not those of Ca\,{\sc iii}, we explicitly write 
[Ca\,{\sc  ii}/H], Ti/Ca\,{\sc ii}, Mn/Ca\,{\sc ii}, [Ti/Ca\,{\sc ii}], and 
[Mn/Ca\,{\sc ii}] in the following. 

Although statistics do not necessarily exclude a more continuous distribution, 
our study indicates that there are two distinct populations of sub-DLAs/DLAs with 
either high or low Ti/Ca\,{\sc ii} ratios separated at 
$[\mathrm{Ti}/$Ca\,{\sc ii}$]\simeq1.0$, reflecting the local dust 
depletion properties. We find that Ca generally tends to deplete severely in
neutral gas that contains Ca and other metals while strong Ti depletion is present
only in relatively dust-rich environments. The observed trends suggest that absorbers 
with high Ti/Ca\,{\sc ii} ratios have dust-to-gas ratios that are substantially 
lower than in the Milky Way and that the observed Ti/Ca\,{\sc ii} ratio represents 
a useful measure for the local dust-to-gas ratio in QSO absorbers.

In our initial survey \citep{bguber} we have studied integrated 
(averaged) Ti/Ca\,{\sc ii} ratios in a large sample of low- and high-redshift 
absorbers because most of these systems exhibit three or less individual absorption 
components in Ca and Ti. %that trace the most dense regions in the absorbers.
Here, we continue our systematic analysis of Ti/Ca\,{\sc ii} ratios in DLAs and 
sub-DLAs by studying intrinsic variations of Ti/Ca\,{\sc ii} in a single 
DLA system at $z=0.313$ toward PKS\,1127$-$145 (J113007$-$144927)  
that is characterized by at least six individual (well-separated) 
velocity components in Ca\,{\sc ii} and Ti\,{\sc ii}. 

In the following, we label the overall DLA system (which consists of at 
least six components) with the term ``DLA'', while we write ``component of 
the DLA'' when we address one of its components. 

The primary goal of our study is to investigate (and separate 
by their dust content) the different neutral gas phases in the overall galaxy 
environment that give rise to this particular absorption system.
Next to Ti, Ca, and Na, we also study intrinsic variations of Mn absorption in this
DLA and compare the trends for Mn/Ca\,{\sc ii} with other DLAs at low- and 
high-redshift.

This paper is structured as follows. In Sect.\,\ref{thedla}, we briefly 
summarize the previously known 
properties of the DLA at $z=0.313$ toward PKS\,1127$-$145, as derived from 
the many previous studies of this absorber. In Sect.\,\ref{data3}, we outline the 
data reduction of our VLT/UVES data and present the analysis method.
In Sect.\,\ref{origin}, we discuss the multiple origins of the DLA sub-components 
in the light of their dust depletion properties.	
A more general discussion of the use of the Mn/Ca\,{\sc ii} ratio as potential 
dust indicator in DLAs and sub-DLAs is provided in Sect.\,\ref{mnca}. Finally, we 
summarize our study in Sect.\,\ref{conclusions}.

\section{The DLA at $z=0.313$ toward PKS\,1127$-$145}
\label{thedla}

The DLA at $z=0.313$ toward the quasar PKS\,1127$-$145 ($z_{\rm em}=1.18$) belongs to 
the best-studied low-redshift DLAs in the literature (e.g., \citealt{bbergeron}; 
\citealt{bguillemin}; \citealt{blane};
\citealt{brao2}; \citealt{bchengalur}; \citealt{bnestor}; \citealt{brao}; 
\citealt{bchen}; \citealt{bchun}; \citealt{bnarayanan}; \citealt{bkanekar};
\citealt{bkacprzaka}; \citealt{bkacprzakb}; \citealt{brichter}; \citealt{bguber}).
There is a very rich data set for this DLA based on various different observatories
including high-resolution, high S/N optical spectra from VLT/UVES, low-resolution 
UV spectral data from HST/STIS and HST/FOS, H\,{\sc i} 21\,cm absorption spectra, 
as well as deep optical imaging (e.g., from HST/WFPC-2) and spectral data for the 
nearby galaxy population.

The DLA at $z=0.313$ toward PKS\,1127$-$145 is characterized by a large neutral hydrogen
column density ($\log N($H\,{\sc i}$)=21.71$; \citealt{blane}; \citealt{brao2}) and a 
complex pattern of at least six individual absorption components in Mg\,{\sc ii} that 
span a velocity range of $\sim 380$\,km\,s$^{-1}$ (\citealt{bkacprzaka}; \citealt{brichter}),
a range that is far too large to be explained by a simple rotating gas disk. 
From the extensive study of \citet{bkacprzakb} follows that the DLA is connected
with a group of at last five galaxies at $z_\mathrm{abs}=0.3$ with $B$-band luminosities 
in the range $(0.04-0.63)\,L_B^{\star}$ and sightline impact parameters 
$\rho/\mathrm{\,kpc}\in[17-241]$. However, it remains unclear which of these 
galaxies (if any) is responsible for the observed strong DLA. \citet{bkacprzaka} 
favor a scenario in which the absorption is caused by tidal debris from a recent
merger event in the galaxy group.

The overall metallicity of the DLA has been constrained to $\sim$0.13 solar
based on the Zn\,{\sc ii}/H\,{\sc i} ratio in the gas \citep{bkanekar}.
Interestingly, this abundance is four to ten times lower than the metallicity of the 
five candidate host galaxies \citep{bkacprzakb}. 
If one takes the observed radial metallicity gradients in DLAs and galaxies 
into account (e.g., \citealt{christensen14}, \citealt{bpilyugina}, \citealt{bpilyuginb}), 
the low metal abundance in the absorber could be explained by gas that resides in 
the outer regions in one of the five candidate galaxies, far away from the central 
star-forming regions (e.g., in the outer gas disk or in a metal-poor accretion flow).  
Alternatively, the DLA system might be connected with a low-surface brightness (LSB) 
galaxy, as suggested by \citet{brao} to explain several low surface-brightness 
emission features close to the quasar. The various H\,{\sc i} 21\,cm absorption 
features detected in
the DLA \citep{bchengalur} are co-aligned in velocity with the Ca\,{\sc ii}
absorption \citep{brichter} tracing the cold neutral gas component in the 
absorber. Yet, the average spin temperature derived from the 21\,cm data is 
$T_{\rm spin}=(910\pm160)$\,K, thus relatively high compared to other DLAs
with large velocity spreads (see survey by \citealt{bchengalur}). 
This possibly suggests that much of the H\,{\sc i} is warm,
as expected for tidal streams (e.g., \citealt{bbruns}).

In this paper, we further investigate the nature of the gas in the DLA by considering
the dust depletion properties in the individual velocity components.

\section{Data origin and analysis method}
\label{data3}
The high resolution data for PKS\,1127$-$145 were taken with 
the Ultraviolet and Visual Echelle Spectrograph (UVES) mounted on the 8.2\,m 
Very Large Telescope (VLT) at Paranal, Chile \citep{Dekker2000}.

The slit width of 1.0 arcsec and a charge-coupled device (CCD) readout with 2$\times$2 binning used for 
all the observations resulted in a spectral resolution power R $\approx$ 
 $45\mathrm{,}000$, corresponding to $\sim$6.7\,km\,s$^{-1}$,
while the $S/N$ is $\approx$100 at $3934$\,\AA\, near the redshifted Ca\,{\sc ii} 
absorption.

To cover the wavelength range of the (redshifted) Na\,{\sc i} D1 and D2 lines as well,
we combined data of three different observation programs publicly available in the 
ESO archive\footnote{67.A-0567(A), 69.A-0371(A), 076.A-0860(A)}.
The details of the observational campaigns are presented in 
Table \ref{table:uves-obs}.

The Common Pipeline Language (CPL version 6.3) of the UVES pipeline was 
used to bias correct and flat field the exposures and then to extract
the wavelength calibrated flux. After the standard reduction, the custom 
software UVES popler\footnote{\url{http://astronomy.swin.edu.au/~mmurphy/UVES_popler/}} 
(version 0.66) was used to combine the extracted echelle orders into 
single 1D spectra. The continuum was fitted with low-order polynomials.

The other UVES data used in Sect.\,5 are taken from our previous survey 
papers (\citet{brichter}; \citet{bguber}). These raw data were reduced as 
part of the SQUAD project (PI: M. Murphy) with a comparable S/N and resolution 
as for PKS\,1127$-$145.

In Fig.\,\ref{linien},
we show the velocity profiles of several transitions from Ca\,{\sc ii}, 
Ti\,{\sc ii}, and Mn\,{\sc ii}. In part due to a lack of spectral coverage,
no diffuse interstellar bands (DIBs) associated with the DLA could be identified
by us in the UVES spectra.

For the spectral analysis, we used the \texttt{fitlyman}-tool in MIDAS to fit 
Voigt-profiles to the individual velocity components of Ca\,{\sc ii} and 
created a component template. 
We then fixed the redshifts of the individual components and, using that template, 
fitted Voigt-profiles also to the other ions considered here (Ti\,{\sc ii}, Mn\,{\sc ii}, 
and Na\,{\sc i}). 
For Na\,{\sc i}, we allowed \texttt{fitlyman} to vary the component's centroid velocity
within 2\,km\,s$^{-1}$ from the fixed ones of the other ions to account for 
the possibility that Na\,{\sc i} traces a slightly different (more dense) 
gas phase than the other above mentioned ions
(see Table \ref{overview}).
The Doppler parameters $b$ were allowed to vary independently for each 
fitted ion, but were fixed for all transitions of the same ion.
Table \ref{line data} in the Appendix lists the vacuum-wavelengths and 
oscillator strengths of the transitions that were fitted by us.
To roughly estimate the H\,{\sc i} column densities in each component, 
we adopted the overall H\,{\sc i} 
column density of the DLA of $\log N($H\,{\sc i}$)=21.71$ (\citealt{blane}; 
\citealt{brao}) and 
calculated the fractional abundance of H\,{\sc i} per component from the observed
Ca\,{\sc ii} column density pattern (i.e., we assume a constant Ca\,{\sc ii}/H\,{\sc i} 
ratio throughout the DLA as a first-order approximate).

We presupposed solar relative abundances given in \citet{basplund} and a standard 
$\Lambda\mathrm{CDM}$-cosmology with 
$H_0=70\,\mathrm{km}\,\mathrm{s}^{-1}\,\mathrm{Mpc}^{-1}$, $\Omega_\mathrm{M}=0.3$, 
and $\Omega_\Lambda=0.7$ in accordance with \citet{bkacprzaka}.
All column densities $N [\mathrm{cm}^{-2}]$ are given in logarithmic units.

\begin{table*}
 \centering
  \caption{Results from the Voigt-profile fitting of the six components in the DLA at 
$z_\mathrm{abs}=0.313$ toward PKS\,1127$-$145. The velocities are given in the 
$z=0.312710$ restframe, as assumed by \citet{bkacprzaka}. 
For Na\,{\sc i}, we allowed \texttt{fitlyman} to vary the component's centroid velocity 
within 2\,km\,s$^{-1}$ from the fixed values for the other ions. The values in brackets 
in column two are these slightly deviant velocities of the components in Na\,{\sc i}. 
The upper limits for $N$(Na\,{\sc i}) were derived from the signal-to-noise ratio 
without the need for specific component-velocities. The Doppler-parameters $b$ are 
from the fits of the Ca\,{\sc ii} lines. For the H\,{\sc i} column densities, a 
proportionality between the column densities of Ca\,{\sc ii}- \& H\,{\sc i} and an 
overall H\,{\sc i} column density of $\log N($H\,{\sc i}$)=21.71$ are assumed.}
  \label{overview}
  \begin{tabular}{lrrrrrrr}
\toprule
Comp.	&$v_\mathrm{rest}$	&$b$\,({Ca\,{\sc ii}})	 &$\log N($H\,\sc{i}$)$	&$\log N($Ca\,{\sc ii}$)$	&$\log N($Ti\,{\sc ii}$)$		&$\log N($Mn\,{\sc ii}$)$ &$\log N($Na\,\sc{i}$)$\\
	&$\left[\mathrm{km}\,\mathrm{s}^{-1}\right]$	&$\left[\mathrm{km}\,\mathrm{s}^{-1}\right]$	&&&&&\\
\midrule
1	&$-46.5$	&11.2	&20.97	&11.85$\pm$0.01	&11.96$\pm$0.03	&12.44$\pm$0.02	&<11.21\\
2	&$-12.3\,(-11.25)$	&6.8	&21.04	&11.92$\pm$0.01	&12.37$\pm$0.02	&12.72$\pm$0.01	&12.28$\pm$0.07\\
3	&$+4.0\,(+2.31)$		&2.6	&20.99	&11.87$\pm$0.01	&(10.12$\pm$1.08)$^a$	&11.09$\pm$0.16	&11.93$\pm$0.04\\
4	&$+15.6\,(+15.88)$	&5.8	&20.88	&11.76$\pm$0.02	&11.78$\pm$0.07	&12.16$\pm$0.03	&11.89$\pm$0.05\\
5	&$+35.3$	&10.6	&21.17	&12.05$\pm$0.01	&12.07$\pm$0.03	&12.56$\pm$0.01	&<11.21\\
6	&$+64.4$	&6.8	&21.14	&12.02$\pm$0.01	&11.84$\pm$0.04	&12.09$\pm$0.02	&<11.21\\
\midrule
total	&$-$	&$-$	&21.71	&12.59$\pm$0.02	&12.76$\pm$0.04	&13.16$\pm$0.02	&12.55$^{+0.11}_{-0.06}$\\
\bottomrule
\end{tabular}
\flushleft
\medskip
\textbf{Notes.} (a) Only weakly constrained; no impact on total budget. 
\end{table*}
%%%%%%%%%%%%%%%%%%%%%%%%%%%%%%%%%%%%%%%%%%%%%%%%%%%%%%%%%%%%%%%%%%%%%%%%%%%%%
%%%%%%%%%%%%%%%%%%%%%%%%%%%%%%%%%%%%%%%%%%%%%%%%%%%%%%%%%%%%%%%%%%%%%%%%%%%%%

\section{Origin of gas in the DLA at $z_\mathrm{abs}=0.313$ toward 
PKS\,1127$-$145} 
\label{origin}

\subsection{Component structure and galaxy environment}

%%%%%%%%%%%%%%%%%%%%%%%%%%%%%%%%%%%%%%%%%%%%%%%%%%%%%%%%%%%%%%%%%%%%%%%%%%%%%
\begin{figure}
\resizebox{\hsize}{!}{\includegraphics{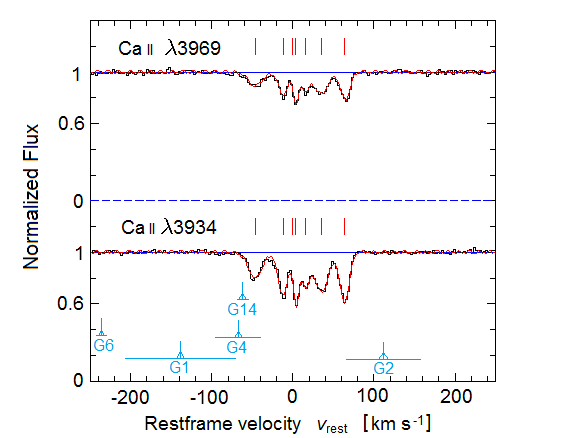}}
\caption{Velocity profiles of Ca\,{\sc ii} $\lambda 3934,3969$ absorption in 
the DLA at $z_\mathrm{abs}=0.312710$ toward PKS\,1127$-$145. The velocities 
of the five galaxies belonging to the galaxy-group at $z_\mathrm{abs}=0.313$ 
along the PKS\,1127$-$145 sightline \citep{bkacprzakb} are indicated in blue. 
The horizontal lines indicate the velocity range allowed for the individual
galaxies based on a simple disk-rotation scheme for these galaxies 
\citep{bkacprzaka}.}
\label{galaxies}
\end{figure}

%%%%%%%%%%%%%%%%%%%%%%%%%%%%%%%%%%%%%%%%%%%%%%%%%%%%%%%%%%%%%%%%%%%%%%%%%%%%%

\begin{figure}
\resizebox{\hsize}{!}{\includegraphics{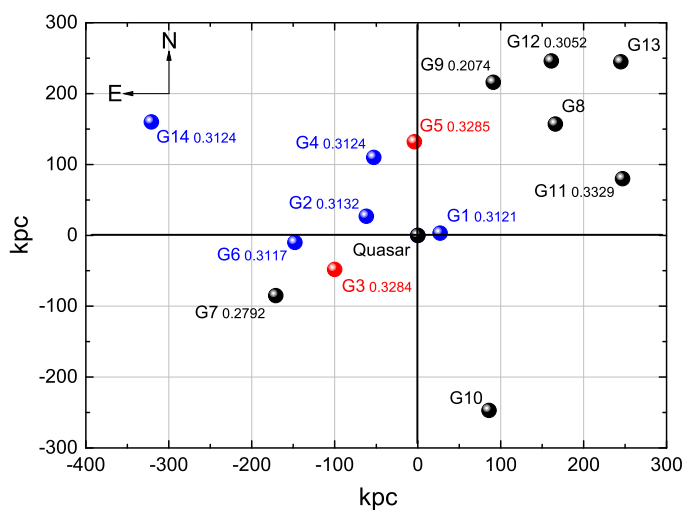}}
\caption{Absorber-galaxy connection for the DLA toward PKS\,1127$-$145 at 
$z_\mathrm{abs}=0.313$ (adopted from \citealt{bkacprzakb}).}
\label{fieldview}
\end{figure}

In Fig.\,\ref{galaxies}, we show the Ca\,{\sc ii} velocity-component structure 
of the DLA toward PKS\,1127$-$145 as seen in our UVES data together 
with the best Voigt-profile fit. 
Six absorption components are identified (see Table \ref{overview}). 
In the lower panel of Fig.\,\ref{galaxies}, we also show the velocities of nearby 
galaxies (G), as identified in the
detailed study of \citet{bkacprzaka}.
In this figure, we indicate the velocity range allowed for the individual
galaxies based on a simple disk-rotation scheme for these galaxies 
\citep{bkacprzaka}. 
\citet{bkacprzaka} identify five galaxies belonging to the same group of galaxies 
that all have redshifts 
$z_\mathrm{abs}\approx 0.313$. Their galaxy ``G1'' is the one with the smallest 
angular distance 
to the quasar sightline ($3.81$\,arcsec,  corresponding to $17.4$\,kpc at this redshift).
However, in view of the observed velocity pattern of the galaxies and the absorber,
a clear correlation between these five galaxies and the DLA components is not evident.
The spatial position of the galaxies in the quasar field is sketched
in Fig.\,\ref{fieldview}, based on the galaxy data presented in \citet{bkacprzaka}.

%%%%%%%%%%%%%%%%%%%%%%%%%%%%%%%%%%%%%%%%%%%%%%%%%%%%%%%%%%%%%%%%%%%%%%%%%%%%%%%%%

%%%%%%%%%%%%%%%%%%%%%%%%%%%%%%%%%%%%%%%%%%%%%%%%%%%%%%%%%%%%%%%%%%%%%%%%%%%%%%%%%%%

\subsection{Gas phases abundances in the individual components}

In Table \ref{overview}, we list the column densities and Doppler-parameters $b$ 
for the individual velocity components 
derived from our Voigt-profile fitting. The measured Ca\,{\sc ii} column densities 
lie all within a small range, $\log N\in[11.76, 12.02]$. For Ti\,{\sc ii} and 
Mn\,{\sc ii}, in contrast,
the dispersion is larger, since the column densities in component 3 are 
substantially smaller than in the other components (see Table \ref{overview}). 
We note at this point that each of the absorption components would be a DLA on its 
own if observed as a single absorption system (see H\,{\sc i} column densities for 
the individual components in Table \ref{overview}, fourth column).

%%%%%%%%%%%%%%%%%%%%%%%%%%%%%%%%%%%%%%%%%%%%%%%%%%%%%%%%%%%%%%%%%%%%%%%%%%%%%%%%%%%

\begin{figure}
\resizebox{\hsize}{!}{\includegraphics{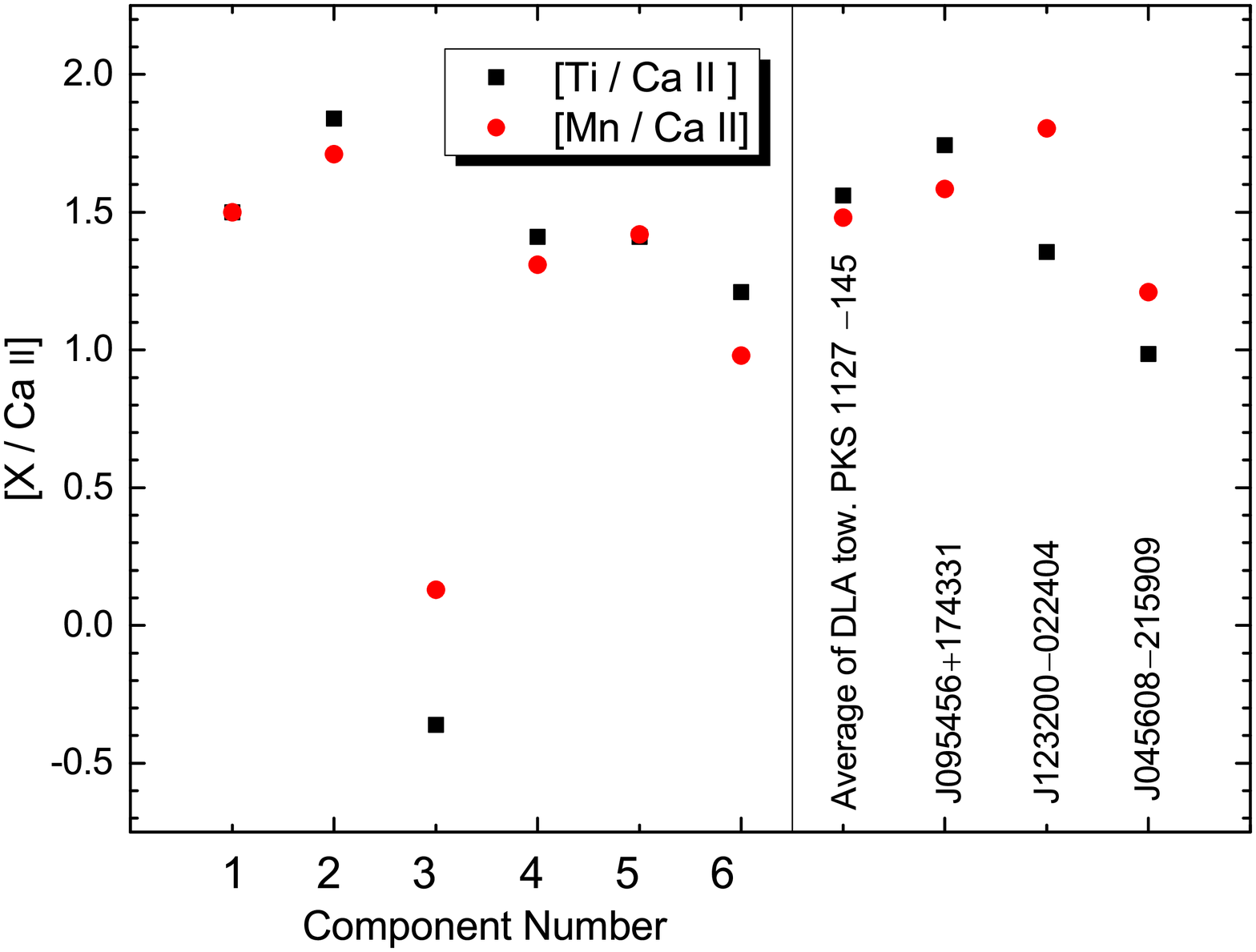}}
\caption{Gas-phase abundances $[$X/Ca\,{\sc ii}$]$ for X$=$Ti and X$=$Mn for the 
six components in the DLA toward PKS\,1127$-$145 and other DLAs and sub-DLAs from 
the \citet{bguber} absorber sample (see also Table \ref{othersystems}).}
\label{3}
\end{figure}

%%%%%%%%%%%%%%%%%%%%%%%%%%%%%%%%%%%%%%%%%%%%%%%%%%%%%%%%%%%%%%%%%%%%%%%%%%%%%%%%%%%

In Fig.\,\ref{3}, we plot the abundance ratios 
%[Ca\,{\sc ii}/H] (assumed to be constant; see previous section), 
[Ti/Ca\,{\sc ii}] and [Mn/Ca\,{\sc ii}] based on
the adopted column densities 
for %H\,{\sc i}, 
Ca\,{\sc ii}, Ti\,{\sc ii}, and Mn\,{\sc ii}.
The plot shows a striking underabundance of Ti and Mn in component 3,
while their abundance is uniform (at a level of [X/Ca\,{\sc ii}$]\approx +1.5$) 
throughout the other components.
The Ca abundance is low overall, $[$Ca\,{\sc ii}$/$H$]\approx -3.5$. 

Since the overall metal abundance in the DLA has been constrained to $\sim 0.13$ solar
(see Sect.\,\ref{thedla}), the observed abundance pattern in Fig.\,\ref{3} 
suggests that all three elements
are depleted into dust grains. 
We note that possible ionization effects, if relevant at these relatively high neutral 
gas columns, would affect only Ca\,{\sc ii} and Mn\,{\sc ii},
but not Ti\,{\sc ii}, because Ti\,{\sc ii} and H\,{\sc i} have identical 
ionization potentials (e.g., \citealt{bmorton}). The depletion of Ca, Ti, 
and Mn appears to be $>2$\,dex in component 3,
suggesting that here the dust-to-gas ratio is significantly higher than
in the other components.
Interestingly, component 3 is also the component with the smallest Doppler parameter, 
$b=2.6$\,km\,s$^{-1}$ (Table \ref{overview}), indicating that the velocity 
dispersion of the gas in component 3 is small.

Only the second, third, and fourth components show Na\,{\sc i} absorption 
(Fig.\,\ref{na1}). Na\,{\sc i} traces cold gas in DLAs at relatively
large gas densities (see discussion in \citealt{brichter}) and therefore the 
detection of Na\,{\sc i} possibly indicate an interstellar origin for these 
components. Although component 3 stands out because of it high dust-to-gas 
ratio and its narrow absorption, it is not the strongest in Na\,{\sc i}, as 
one might expect intuitively. However, also Na possibly is depleted into dust 
grains in such dense interstellar environments, which complicates the 
interpretation of the observed Na\,{\sc i} line strengths.

Component 3, which is narrow and strong in Ca\,{\sc ii}, weak in Ti\,{\sc ii} 
and Mn\,{\sc ii}, shows all spectral signatures expected for a gaseous disk 
of a spiral galaxy.
This is because in such an environment, one would expect cool, dense gas 
with a small velocity 
dispersion and a relatively high dust-to-gas ratio. 
Since none of the clearly detected 
galaxies presented in \citet{bkacprzaka} lies very close ($\rho <10$\,kpc) to the
quasar sightline or is close in velocity (Fig.\,\ref{galaxies}), and because 
the metallicity of the DLA
is lower than that of the surrounding galaxy population (see Sect.\,\ref{thedla}), 
the galaxy disk that gives rise to the absorption in component 3 is possibly related
to a faint LSB galaxy that lies in front of PKS\,1127$-$145. 
This scenario was also preferred by \citet{brao} to explain several
low surface-brightness emission features close to the quasar.

The broader and less dust-rich satellite components, in contrast, could well be related 
to relatively warm, diffuse H\,{\sc i} gas from tidal 
debris due to the interaction between 
the galaxy-group members, as proposed by \citet{bkacprzakb}. 
The mild depletion values for Ca and Ti ($<2$ dex) in these components are 
very similar to those found in the Magellanic Stream \citep{brichter2013},
the most nearby tidal gas stream in the Milky Way halo.

\subsection{Comparison with other DLAs}

In our previous survey of intervening Ti\,{\sc ii}/Ca\,{\sc ii} absorbers at 
low redshift \citep{bguber}, we concluded that the observed abundance of 
Ca in DLAs and sub-DLAs depends only relatively little on the local metallicity 
and dust content. Because of the high condensation temperature of Ca 
($T_\mathrm{c}>1.5\times10^{3}\,\mathrm{K}$)\footnote{For a specific element, 
$T_\mathrm{C}$ is the temperature below which 50\,\% of the total amount is 
condensed into the dust phase \citep{bwasson}.}, this element is incorporated 
into dust grains very efficiently, even at low metallicities and 
low dust-to-gas ratios.
The depletion of Ti in DLAs, in contrast, is much lower in warm, diffuse systems with 
low-metallicity gas, while it increases dramatically in dense regions with high 
metallicities.
The observed trends for Ti and Ca seen in other absorption-line systems 
(see compilation by \citealt{bguber} and references therein)
lead us to the conclusion that the Ti/Ca\,{\sc ii} ratio in DLAs serves as an 
excellent tracer for the local dust-to-gas ratio.

A look at Fig.\,\ref{3} immediately shows that the third component of our DLA
has a Ti/Ca\,{\sc ii} ratio that is substantially lower than in the other 
absorption components.
Indeed, for component 3, we derive $[\mathrm{Ti}/$Ca\,{\sc ii}$]=-0.36$, while for
the other components, we find $[\mathrm{Ti}/$Ca\,{\sc ii}$]>1.4$.
For Mn\,{\sc ii}, the trend is similar as for Ti\,{\sc ii}, which will be further
discussed in Sect.\,\ref{mnca}.
With $[\mathrm{Ti}/$Ca\,{\sc ii}$]=-0.36$, component 3 shows a depletion
characteristics that is very similar to the Milky Way disk \citep{bwelty},
while the other components have Ti/Ca\,{\sc ii} ratios similar to those found
in the SMC, LMC, and in other low-redshift DLAs with low metallicities \citep{bguber}.

\section{Mn/Ca\,{\sc ii} as indicator for the dust-to-gas ratio in DLAs}
\label{mnca}

Figure\,\,\ref{3} shows that the depletion of Mn  follows that of Ti very 
closely in the DLA toward PKS\,1127$-$145. In addition, the 
Mn/Ca\,{\sc ii} ratio shows a trend that 
is very similar to that for Ti/Ca\,{\sc ii}. The question arises whether 
this trend is typical for DLAs and sub-DLAs and
whether the Mn/Ca\,{\sc ii} ratio could be used as a dust indicator
for systems where no Ti information is available.

To further study this interesting aspect, we reinvestigated the Ca\,{\sc ii} 
absorbers presented in \citet{brichter} and analyzed Mn\,{\sc ii} absorption in
these systems. The results are presented in Table \ref{othersystems}.
Only those systems are listed for which a reliable value for log $N$(Mn\,{\sc ii}) 
or an upper limit could be derived.

%%%%%%%%%%%%%%%%%%%%%%%%%%%%%%%%%%%%%%%%%%%%%%%%%%%%%%%%%%%%%%%%%%%%%%%%%%%%%%%%%%

\begin{figure}
\resizebox{\hsize}{!}{\includegraphics{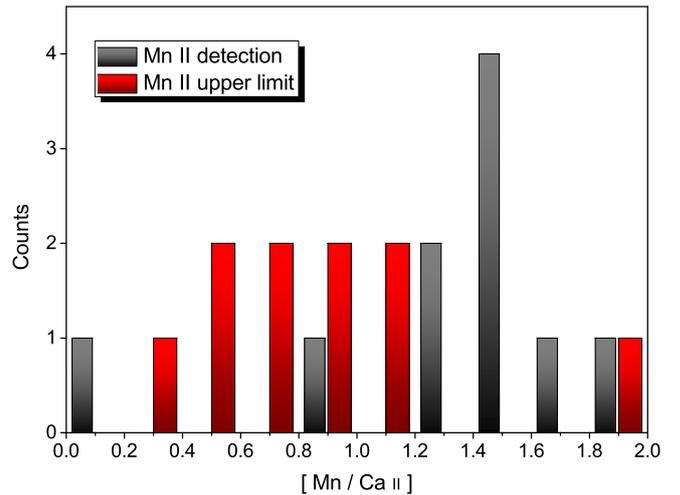}}
\caption{Mn/Ca\,{\sc ii} ratios in the six components in the DLA 
toward PKS\,1127$-$145 and the other DLAs and sub-DLAs studied in \citet{bguber}.}
\label{histogram}
\end{figure}

%%%%%%%%%%%%%%%%%%%%%%%%%%%%%%%%%%%%%%%%%%%%%%%%%%%%%%%%%%%%%%%%%%%%%%%%%%%%%%%%%

Generally, the Mn gas phase abundances in these DLAs and sub-DLAs, indeed, 
seem to closely trace the abundances of Ti, as the three additional Ca\,{\sc ii} 
systems with Mn\,{\sc ii} detections show.
The resulting histogram of $[\mathrm{Mn}/$Ca\,{\sc ii}$]$-values 
(Fig.\,\ref{histogram}), which includes
the absorption components of the DLA toward PSK\,1127$-$145 at $z_\mathrm{abs}=0.313$
as individual data points, looks very similar to 
the histogram for $[\mathrm{Ti}/$Ca\,{\sc ii}$]$, as presented in \citet{bguber}.
The Mn\,{\sc ii} upper limit systems mostly have $[\mathrm{Mn}/$Ca\,{\sc ii}$]<1.2$ 
while the systems with a measured value for $N($Mn\,{\sc ii}$)$ mostly have larger 
$[\mathrm{Mn}/$Ca\,{\sc ii}$]$ ratios.
Therefore, similar as for Ti, there seems to be a separation 
into two classes of absorbers: 

\begin{enumerate}
\item Class 1: high Mn/Ca\,{\sc ii} ratios with $[\mathrm{Mn}/$Ca\,{\sc ii}$]>0.8$ and
\item Class 2: low Mn/Ca\,{\sc ii} ratios with $[\mathrm{Mn}/$Ca\,{\sc ii}$]<0.8$.
\end{enumerate}

As for Ti/Ca\,{\sc ii} systems, Class 1 Mn/Ca\,{\sc ii} absorbers 
trace gas with relatively little
dust (Mn only mildly depleted, Ca strongly depleted) while Class 2 systems trace 
gas with a relatively high dust-to-gas ratio (Mn strongly depleted, similar to Ca).

%%%%%%%%%%%%%%%%%%%%%%%%%%%%%%%%%%%%%%%%%%%%%%%%%%%%%%%%%%%%%%%%%%%%%%%%%%%%%%%%%%%

\begin{figure}
\resizebox{\hsize}{!}{\includegraphics{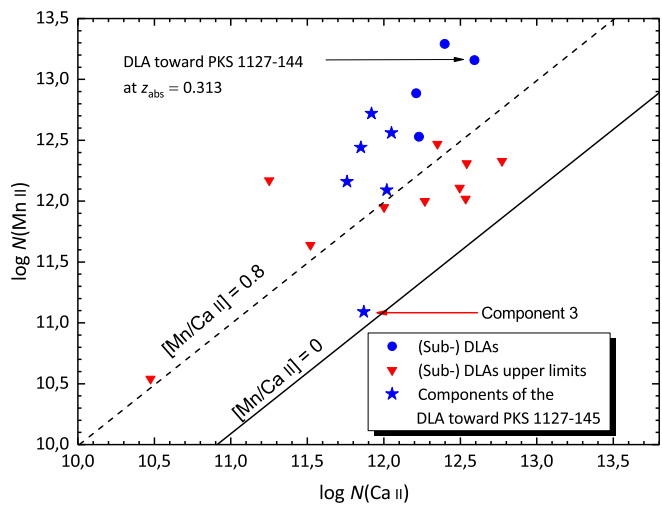}}
\caption{Relation between column densities of Mn\,{\sc ii} and Ca\,{\sc ii} in the 
DLA toward PKS\,1127$-$145 and other quasar absorption line systems. Only in 
component 3 of the DLA toward PKS\,1127$-$145, the depletion of Mn into dust grains 
is as strong as for Ca, so that the resulting [Mn/Ca\,{\sc ii}] ratio is near the 
solar value.}
\label{MnvsCa}
\end{figure}

%%%%%%%%%%%%%%%%%%%%%%%%%%%%%%%%%%%%%%%%%%%%%%%%%%%%%%%%%%%%%%%%%%%%%%%%%%%%%%%%%%%%

Also the relation between the Mn\,{\sc ii} and Ca\,{\sc ii} column densities 
in DLAs and sub-DLAs  
(Fig.\,\ref{MnvsCa}) is very similar to the one derived for Ti\,{\sc ii} 
and Ca\,{\sc ii} (\citet{bguber}; their Fig.\,1).
Most absorption components of the DLA toward PKS\,1127$-$145 follow the 
trend seen in other low-$z$ DLAs while component 3 stands out, having a
nearly solar Mn/Ca\,{\sc ii} ratio.

%%%%%%%%%%%%%%%%%%%%%%%%%%%%%%%%%%%%%%%%%%%%%%%%%%%%%%%%%%%%%%%%%%%%%%%%%%%%%%%%%%%%

\begin{figure}
\resizebox{\hsize}{!}{\includegraphics{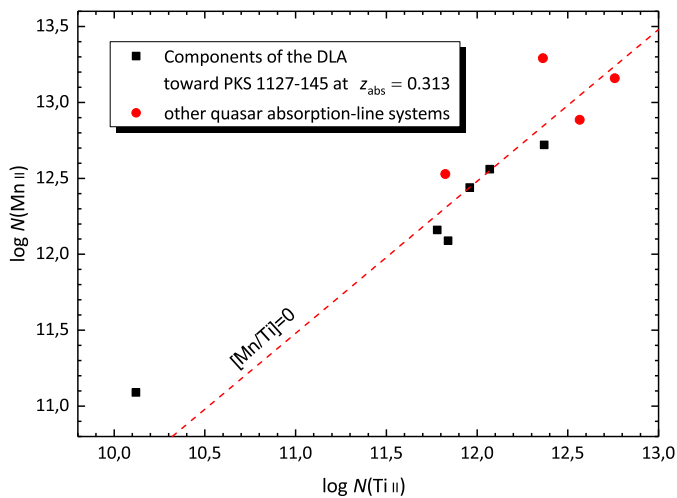}}
\caption{Relation between Mn\,{\sc ii} and Ti\,{\sc ii} for the DLA toward 
PKS\,1127$-$145 and other quasar absorption-line systems.}
\label{MnTi}
\end{figure}
%%%%%%%%%%%%%%%%%%%%%%%%%%%%%%%%%%%%%%%%%%%%%%%%%%%%%%%%%%%%%%%%%%%%%%%%%%%%%%

Beside dust depletion, different nucleosynthetic abundance patterns of 
elements have influence on their gas-phase abundance ratios.
While Ca is an $\upalpha$ element and Ti shows an abundance pattern that is 
similar to the one of an $\upalpha$ element (e.g., \citealt{bwelty}), 
Mn is an iron-peak element.
\citet{bdecia} have addressed nucleosynthetic influences on dust depletion
measurements in detail.
They express that an overabundance of $\upalpha$ elments compared to 
iron-peak elements (short: $\upalpha$ enhancement, $\alpha=[\alpha/\mathrm{Fe}]$) 
is maximal for low-metallicity DLAs, which often have low dust-to-gas-ratios 
\citep{bdecia}. However, possible $\upalpha$ enhancements can be expected to 
be negligible for metal-rich, typically dust-rich, environments, as in 
molecular clouds in the Milky Way.
While \citet{bdecia} expect an averaged overabundance of alpha elements 
compared to Fe of $\alpha_\mathrm{max}\approx +0.35$\,dex for [Zn/Fe$]\approx 0$, 
for Mn, they find an underabundance of $\alpha_\mathrm{max,Mn}-0.39$\,dex.
In fact, in the Class 1 regime, where we expect a comparably low dust-to-gas 
ratio and low metallicities, the different abundance patterns of Mn and Ca 
could decrease the Mn/Ca\,{\sc ii} ratio.
Nevertheless, our observations suggest the opposite, indicating that, 
although the different abundance patterns influence the Mn/Ca\,{\sc ii} ratio, 
the effect of dust depletion on the Mn/Ca\,{\sc ii} ratio is dominating.

The similarity of Mn and Ti as tracers for the dust content in DLAs is further
underlined by the trend seen in in Fig.\,\ref{MnTi}. There is a rather tight relation
between log $N$(Mn\,{\sc ii}) and log $N$(Ti\,{\sc ii}) with a slope that
is consistent with a solar Mn/Ti ratio, indicating that 
both ions are similarly sensitive to the local gas phase metal abundance and
thus similarly sensitive to the local dust-to-gas ratio.

\section{Conclusions}
\label{conclusions}

In this paper, we have studied the gas-phase abundances of Ca, Ti, Mn, and 
Na in the various absorption components of a DLA at $z_\mathrm{abs}=0.313$ 
toward PKS\,1127$-$145 based on VLT/UVES spectra
to explore the dust depletion of these elements and to constrain the origin 
of the absorber and its components.  
From previous measurements it is known that the DLA toward PKS\,1127$-$145 
is characterized by a complex sub-component structure, a large neutral 
gas column density (log $N$(H\,{\sc i}$)=21.71$), and a relatively 
low metallicity $(\sim 0.13\,\mathrm{solar})$.

The following results have been obtained for the DLA:

\begin{enumerate}

\item The UVES data imply that the DLA is composed of six major absorption 
components in Ca, Ti, and Mn, which trace individual neutral gas layers 
within the DLA. These six components span a velocity range of 
$\Delta v \approx 140$\,km\,s$^{-1}$.

\item We provide high-precision Voigt-profile fits to the individual 
absorption components and determine column densities (or limits) and 
Doppler-parameters for Ca\,{\sc ii}, Ti\,{\sc ii}, Mn\,{\sc ii}, and 
Na\,{\sc i}. 
Assuming a proportionality between $N($H\,{\sc i}$)$ and 
$N($Ca\,{\sc ii}$)$, we find that every single sub-component would 
represent a DLA on its own.

\item Our study indicates a striking underabundance of Ti and Mn in 
component 3 compared to the other 5 components. The depletion values 
of Ca, Mn, and Ti in component 3 are $>$2\,dex, suggesting a 
considerably higher dust-to-gas ratio in this absorption component 
compared to the other ones. The three inner components (components 2, 3, and 4) 
are the only ones that show absorption by Na\,{\sc i}, indicating 
the presence for colder and denser gas.

\item Component 3 shows very narrow absorption with a Doppler-parameter 
$b$ of only $2.6$ km\,s$^{-1}$. It is relatively strong in Ca\,{\sc ii}, 
but weak in Ti\,{\sc ii} and Mn\,{\sc ii}, pointing toward a large 
neutral gas column with spatially confined, dense gas that contains 
substantial amounts of dust. This component therefore shows all 
signatures of a gaseous disk of a spiral galaxy. These spectral 
signatures speak for a faint LSB galaxy in front of PKS\,1127$-$145, 
as the origin for this third component, a scenario that is also 
preferred by the previous study of \citet{brao}.

\item The other, broader and less dust-rich, outer components probably 
are related to warm, diffuse gas that stems from tidal debris caused 
by the interaction between the various galaxy-group members that are 
associated with the DLA, as proposed by \citet{bkacprzakb}. The 
relatively mild depletion values for Ca and Ti are very similar to 
those found in the Magellanic Stream \citep{brichter2013}, the most 
nearby tidal gas stream in the Milky Way halo.

\item With $[\mathrm{Ti}/$Ca\,{\sc ii}$]=-0.36$, component 3 shows 
depletion characteristics similar to those in the Milky Way disk 
\citep{bwelty} while the other components have Ti/Ca\,{\sc ii} 
ratios similar to those found in the SMC, LMC, and in other DLAs 
\citep{bguber}.

\end{enumerate}

From the comparison between the DLA toward PKS\,1127$-$145 and other 
Ca, Ti, and Mn absorbing environments we draw the following, additional 
conclusions: 

\begin{enumerate}

\item A more detailed look into the Mn absorption properties in DLAs 
suggests that, generally, the Mn\,{\sc ii} gas-phase abundance seems 
to be closely related to the gas-phase abundance of Ti\,{\sc ii}. 
Following the classification scheme for Ti/Ca\,{\sc ii} absorption 
in intervening absorbers presented in \citet{bguber}, we suggest 
that sub-DLAs and DLAs can be subdivided into two classes according 
to their Mn/Ca\,{\sc ii} ratio:

\begin{itemize}
\item Class 1: high Mn/Ca\,{\sc ii} ratios with $[\mathrm{Mn}/$Ca\,{\sc ii}$]>0.8$ and
\item Class 2: low Mn/Ca\,{\sc ii} ratios with $[\mathrm{Mn}/$Ca\,{\sc ii}$]<0.8$.
\end{itemize} 

\item The tight relation between $\log N($Mn\,{\sc ii}$)$ and 
$\log N($Ti\,{\sc ii}$)$ indicates that both ions are similarly 
sensitive to the local gas-phase metal abundance and thus similarly 
sensitive to the local dust-to-gas ratio. Therefore, the determination 
of precise abundances of Ca, Ti, and Mn in low redshift absorption-line systems 
from optical observations alone may provide important information on 
the distribution and cross section of dust-rich environments inside and 
outside of galaxies.

\end{enumerate}

\begin{acknowledgements}

Based on observations made with ESO Telescopes at the La Silla Paranal 
Observatory under programs 67.A-0567(A), 69.A-0371(A), 076.A-0860(A).

\end{acknowledgements}

%%%%%%%%%%%%%%%%%%%%%%%%%%%%%%%%%%%%%%%%%%%%%%%%%%%%%%%%%%%%%%%%%%%%%%%%%%%%%
%%%%%%%%%%%%%%%%%%%%%%%%%%%%%%%%%%%%%%%%%%%%%%%%%%%%%%%%%%%%%%%%%%%%%%%%%%%%%
\appendix
\section{Additional figures and tables}
%%%%%%%%%%%%%%%%%%%%%%%%%%%%%%%%%%%%%%%%%%%%%%%%%%%%%%%%%%%%%%%%%%%%%%%%%%%%%

\begin{figure}
\resizebox{\hsize}{!}{\includegraphics{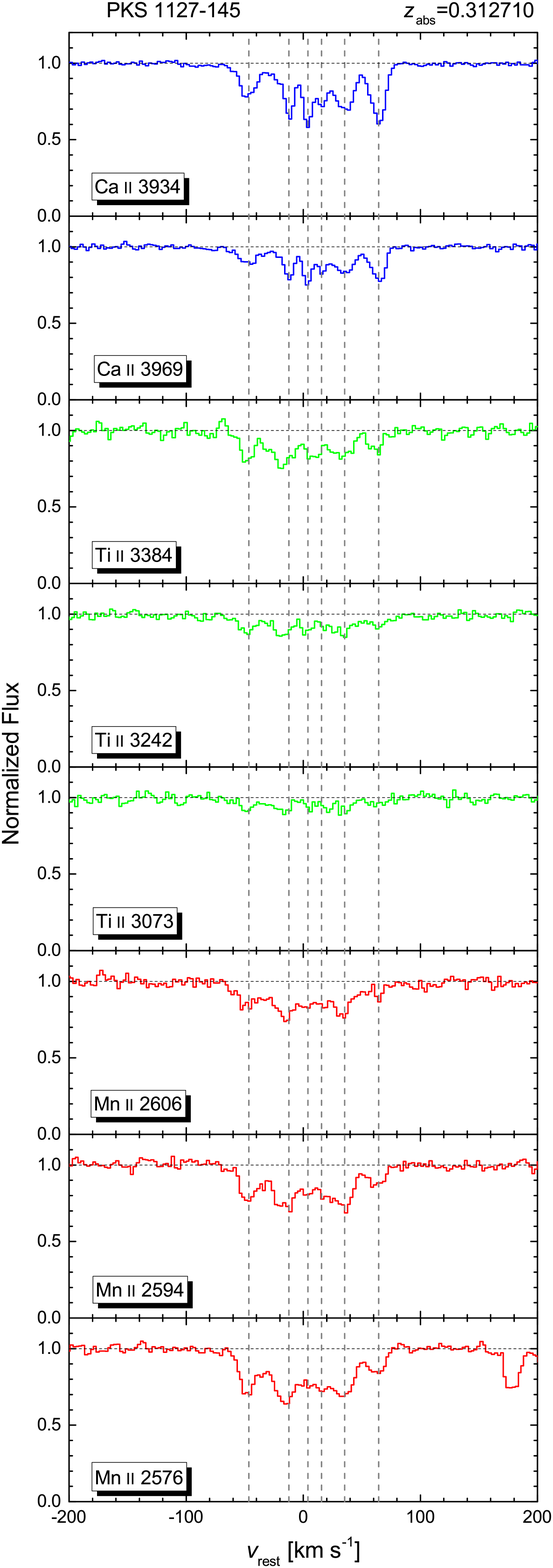}}
\caption{Velocity profiles for various transitions of Ca\,{\sc ii}, 
Ti\,{\sc ii}, and Mn\,{\sc ii} in the $z_\mathrm{abs}=0.313$ DLA toward 
PKS\,1127$-$145 based on archival VLT/UVES data.}
\label{linien}
\end{figure}

%%%%%%%%%%%%%%%%%%%%%%%%%%%%%%%%%%%%%%%%%%%%%%%%%%%%%%%%%%%%%%%%%%%%%%%%%%%%

\begin{table*}
\centering
\caption{Summary of UVES observations for PKS\,1127$-$145. \label{table:uves-obs}}
\begin{tabular}{lccc}
\toprule
setting $\lambda_c$ (nm)       &  T$_{\rm exp}$ (s)    &Date       & Prog.ID    \\
\midrule
346+580                        &30,600                   &2001-05-01,2 / 2001-06-12,13 & 67.A-0567(A)\\
390+564                        &19,200                   &2002-04-05 & 69.A-0371(A)\\
390+580                        &1,200                    &2006-01-04 & 076.A-0860(A)\\
437+760                        &1,200                    &2006-01-04 & 076.A-0860(A)\\
\bottomrule
\end{tabular}
\vspace*{0,1cm}
\end{table*}

%%%%%%%%%%%%%%%%%%%%%%%%%%%%%%%%%%%%%%%%%%%%%%%%%%%%%%%%%%%%%%%%%%%%%%%%%%%%%

\begin{figure}
\resizebox{\hsize}{!}{\includegraphics{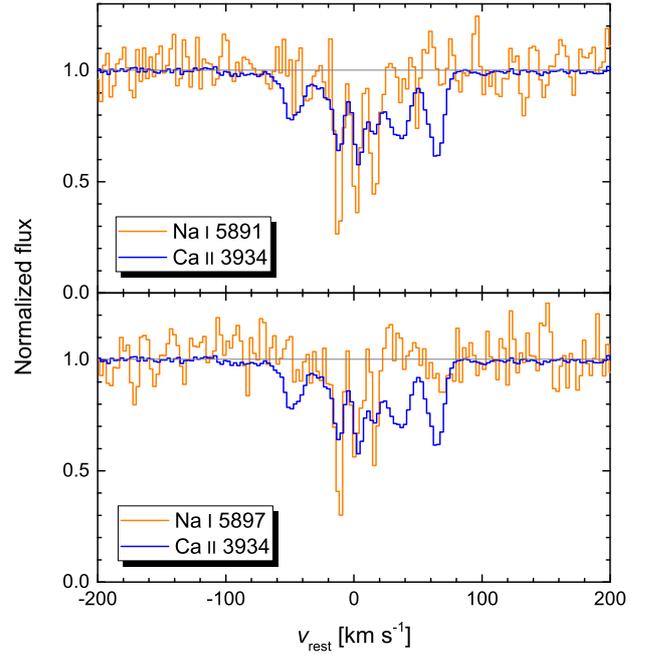}}
\caption{Velocity profiles of Na\,{\sc i} and Ca\,{\sc ii} toward 
PKS\,1127$-$145 at $z_\mathrm{abs}=0.313$. Only in components 2, 3, and 4, 
Na\,{\sc i} absorption is detected.}
\label{na1}
\end{figure}

%%%%%%%%%%%%%%%%%%%%%%%%%%%%%%%%%%%%%%%%%%%%%%%%%%%%%%%%%%%%%%%%%%%%%%%%%%%%%%

%%%%%%%%%%%%%%%%%%%%%%%%%%%%%%%%%%%%%%%%%%%%%%%%%%%%%%%%%%%%%%%%%%%%%%%%%%%%%%
% ADDITIONAL FIGURES NOT REQUIRED FOR PAPER, BUT FOR THESIS
%
%\begin{figure}
%  \resizebox{\hsize}{!}{\includegraphics{331.eps}}
%  \caption{Velocity-plots for the DLA PKS\,$0952+179$ at %$z_\mathrm{abs}=0.23782$.}
%  \label{331}
%\end{figure}

%%%%%%%%%%%%%%%%%%%%%%%%%%%%%%%%%%%%%%%%%%%%%%%%%%%%%%%%%%%%%%%%%%%%%%%%%%%%%%

%%%%%%%%%%%%%%%%%%%%%%%%%%%%%%%%%%%%%%%%%%%%%%%%%%%%%%%%%%%%%%%%%%%%%%%%%%%%%%

%\begin{figure}
%  \resizebox{\hsize}{!}{\includegraphics{404.eps}}
%  \caption{Velocity-plots for the DLA toward PKS\,$1229-021$ at 
%$z_\mathrm{abs}=0.39498$.}
%  \label{404}
%\end{figure}

%%%%%%%%%%%%%%%%%%%%%%%%%%%%%%%%%%%%%%%%%%%%%%%%%%%%%%%%%%%%%%%%%%%%%%%%%%%%%

%%%%%%%%%%%%%%%%%%%%%%%%%%%%%%%%%%%%%%%%%%%%%%%%%%%%%%%%%%%%%%%%%%%%%%%%%%%%%

%\begin{figure}
%  \resizebox{\hsize}{!}{\includegraphics{909.eps}}
%  \caption{Velocity-plots for the sub-DLA toward PKS\,$0454-22$ at 
%$z_\mathrm{abs}=0.47439$.}
%  \label{909}
%\end{figure}

%%%%%%%%%%%%%%%%%%%%%%%%%%%%%%%%%%%%%%%%%%%%%%%%%%%%%%%%%%%%%%%%%%%%%%%%%%%%%

\begin{table}
\caption{Vacuum wavelengths and oscillator strengths for the transitions 
used in this paper.}
 \label{line data}
 \centering
 \begin{tabular}{lll}
  \toprule
   Ion \& transition     & $\lambda_\mathrm{vac}$       & $f$\\
                                                &$[\SI{}{\angstrom}]$   & \\
 \midrule
Ca\,{\sc ii} $\lambda3934$ 	& 3934.7750 & 0.6267\\
Ca\,{\sc ii} $\lambda3969$ 	& 3969.5901	& 0.3116\\
Mn\,{\sc ii} $\lambda2576$	& 2576.877	& 0.361\\
Mn\,{\sc ii} $\lambda2594$	& 2594.499  & 0.280\\
Mn\,{\sc ii} $\lambda2606$  & 2606.462  & 0.198\\
Na\,{\sc i}  $\lambda5891$  & 5891.5833 & 0.640\\
Na\,{\sc i}  $\lambda5897$  & 5897.5581 & 0.3201\\
Ti\,{\sc ii} $\lambda3073$ 	& 3073.863 	& 0.121\\
Ti\,{\sc ii} $\lambda3242$ 	& 3242.918 	& 0.232\\
Ti\,{\sc ii} $\lambda3384$ 	& 3384.730 	& 0.358\\
\bottomrule
\end{tabular}
\flushleft
\medskip
\textbf{References.} All data are from \citet{bmorton}.
\end{table}

\begin{table*}
 \centering
  \caption{Column densities of H\,{\sc i}, 
Ca\,{\sc ii}, Ti\,{\sc ii}, and Mn\,{\sc ii} for DLAs and sub-DLAs 
from the sample of \citet{bguber}.}
  \label{othersystems}
  \begin{tabular}{llrrrr}
\toprule
QSO-sightline    &$z_\mathrm{abs}$ &$\log N($H\,{\sc i}$)$        &$\log N($Ca\,{\sc ii}$)$       &$\log N($Ti\,{\sc ii}$)$               &$\log N($Mn\,{\sc ii}$)$\\
\midrule
J095456$+$174331        &0.23782        &21.32$^a$  &12.21$\pm$0.03 &12.57$\pm$0.05 &12.89$\pm$0.05\\
J113007$-$144927$^*$	&0.31271		&21.71$^b$  &12.59$\pm$0.02 &12.76$\pm$0.04 &13.16$\pm$0.02\\
J123200$-$022404 		&0.39498        &20.6$^c$   &12.40$\pm$0.02 &12.36$\pm$0.05 &13.29$\pm$0.02\\
J045608$-$215909        &0.47439        &19.5$^d$   &12.23$\pm$0.02 &11.83$\pm$0.05 &12.53$\pm$0.02\\
\midrule
\multicolumn{6}{c}{Additional Ca\,{\sc{ii}} systems with upper limits for $N($Mn\,\sc{ii}$)$}\\
\midrule
J124646$-$254749        &0.49282        &$-$    &12.77$\pm$0.08 	&$<11.93$        &$<12.33$\\
J220743$-$534633        &0.43720        &$-$    &12.00$\pm$0.07 	&$<11.50$        &$<11.95$\\
J051707$-$441055        &0.42913        &$-$    &10.48$\pm$0.06 	&$<10.20$        &$<10.54$\\
J232820$+$002238        &0.41277        &$-$    &11.52$\pm$0.08  	&$<11.14$        &$<11.64$\\
J224752$-$123719        &0.40968        &$-$    &12.27$\pm$0.04 	&$<11.34$        &$<12.00$\\
J050112$-$015914        &0.40310        &$-$    &12.35$\pm$0.05  	&$<12.08$        &$<12.47$\\
J110325$-$264515        &0.35896        &$-$    &11.25$\pm$0.04 	&$<10.55$        &$<12.17$\\
J142253$-$000149        &0.34468        &$-$    &12.35$\pm$0.05 	&$<11.88$        &$<12.02$\\
J231359$-$370446        &0.33980        &$-$    &12.54$\pm$0.03 	&$<12.19$        &$<12.31$\\
J102837$-$010027        &0.32427        &$-$    &12.50$\pm$0.03 	&$<11.68$        &$<12.11$\\
\bottomrule
\end{tabular}
\medskip
\footnotesize
\flushleft
\textbf{Notes.} $^*$\,This absorption line system is the multicomponent 
DLA toward PKS\,1127$-$144 at $z_\mathrm{abs}=0.31271$ studied in this paper. 
The given column densities for this absorber (for Ca\,{\sc ii}, Ti\,{\sc ii}, 
and Mn\,{\sc ii}), each, are summed over the six  components as given in the 
last row of Table \ref{overview}.\\
\medskip
\textbf{References}. (a) \citet{brao}; (b) \citet{blane}; 
(c) \citet{blebrun}; (d) \citet{bturnshek}.
\end{table*}

\end{document}